\documentclass[english,twocolumn,superscriptaddress,preprintnumbers,amsmath,amssymb,aps,prd,nofootinbib,showpacs,showkeys]{revtex4}
\usepackage{graphicx}
\usepackage{amssymb}
\usepackage{bbm} %for Z_2 font%
\def\citet#1{\cite{#1}}
\makeatletter
\newcommand{\smallfrac}[2]{\mbox{\small ${\displaystyle
      \frac{#1}{#2}}$}}

\def\remove#1{{\em #1}}
\def\comment#1{\fbox{\begin{minipage}{\columnwidth}#1\end{minipage}}}
\def\remove#1{}
\def\comment#1{}

\makeatother

\usepackage{babel}

\begin{document}

\newcommand{\AU}{School of Chemistry \& Physics, The University of
 Adelaide, South Australia 5005, Australia}
\newcommand{\IC}{Department of Physics, Imperial College London,
  Prince Consort Road, London SW7 2AZ, UK}
\newcommand{\MU}{Department of Mathematical Physics, National
  University of Ireland Maynooth, Maynooth, County Kildare, Ireland}    
\newcommand{\TUD}{Institut f\"ur Kernphysik, Technische Universit\"at
  Darmstadt, Schlossgartenstr.~9, 64289 Darmstadt, Germany}

\title{'t Hooft-Polyakov monopoles in lattice SU($N$)+adjoint Higgs theory}

\author{S. Edwards}
\affiliation{\AU}

\author{D. Mehta}
\affiliation{\AU}
\affiliation{\IC}
\affiliation{\MU}

\author{A. Rajantie}
\affiliation{\IC}
%\email{a.rajantie@imperial.ac.uk}

\author{L. \surname{von Smekal}}
\affiliation{\AU}
\affiliation{\TUD}

%\date{\today}
\date{June 29, 2009}

\begin{abstract}
We investigate twisted C-periodic boundary conditions in SU($N$)
gauge field theory with an adjoint Higgs field. We show that with
a suitable twist for even $N$ one can impose a non-zero magnetic charge relative
to residual U(1) gauge groups in the broken phase, thereby creating
a 't~Hooft-Polyakov magnetic monopole. This makes it possible to use
lattice Monte-Carlo simulations to study the properties of these monopoles
in the quantum theory. 
\end{abstract}



\maketitle

\section{Introduction}

't~Hooft-Polyakov monopoles~\citet{'tHooft:1974qc,Polyakov:1974ek}
play an important role in high energy physics, partly because their
existence as physical particles is a general prediction of grand unified
theories, and partly because they provide a way to study
non-perturbative properties of quantum field theories through
electric-magnetic dualities~\citet{Montonen:1977sn}. 
Most of the existing studies of monopoles in non-supersymmetric
theories have been restricted to the level of classical solutions, and
little is known about quantum mechanical effects.
Calculation of even leading-order quantum corrections to solitons
is hard, and can usually only be done in simple one-dimensional models~\citet{Dashen:1974ci}.

Lattice Monte Carlo simulations provide an alternative, fully non-perturbative
approach~\citet{Ciria:1993yx}. However, because of their non-perturbative
nature, they always describe the true ground state of the system and
therefore do not allow one to specify a background about which the
theory is quantised.

There are two general approaches to calculating properties of monopoles
and other solitons in Monte Carlo simulations. One can either define
suitable creation and annihilation operators and measure their correlators~\citet{Veselov:1991pr,DelDebbio:1994sx,Frohlich:1998wq},
or one can impose boundary conditions that restrict the path integral
to a non-trivial topological sector~\citet{Smit:1989vg}. The former
approach is closer in spirit to usual Monte Carlo simulations and
in principle it gives access to a wide range of observables including,
e.g., the vacuum expectation value of the monopole field. However,
because monopoles are surrounded by a spherical infinite-range magnetic
Coulomb field, it is difficult to find a suitable operator and separate
the true ground state from excited states.

Instead, while non-trivial boundary conditions provide access to a
more limited set of observables, they ensure that the monopole is
always in its ground state. Early attempts to simulate monopoles were
based on fixed boundary conditions~\cite{Smit:1993vt}, but this
introduced large finite-size effects. To avoid them, one needs to use
boundary conditions that are periodic up to the symmetries of the
theory. Such boundary conditions were introduced for the SU($2$)
theory in Ref.~\citet{Davis:2000kv}, and they were used to calculate
the mass of the monopoles in
Refs.~\citet{Davis:2001mg,Rajantie:2005hi}. 

In this paper, we generalise this result to SU($N$) gauge group with
$N>2$. This is important for several reasons. Many analytical results
are only valid in the large-$N$ limit, and for grand unified theory
monopoles one needs SU($5$) or larger groups. The SU($2$) group is also
somewhat special, and a richer theoretical structure with new questions
arises when one goes beyond it. For instance, there can be several
different monopole species and unbroken non-Abelian gauge groups.

We find that as in the SU($2$) theory, monopoles can be created by boundary
conditions that consist of complex conjugation and a topological non-trivial
gauge transformation, but only for even $N$. The boundary conditions
treat all monopole species in the same way, so we cannot single out
one for creation. Instead of actually fixing the magnetic charge,
we can only choose between odd and even charges. However, even with
these limitations, the boundary conditions make it possible to measure
the monopole mass.

The paper is organsied as follows. In Sections~\ref{sec:continuum}
and~\ref{sec:lattice}, we review the definitions of the magnetic
field and magnetic charge in the SU($N$)+adjoint Higgs theory in
the continuum and on the lattice, respectively. In Section~\ref{sec:mass}
we show how the monopole mass is expressed in terms of partion functions
for different topological sectors. In Section~\ref{sec:twist}, we
introduce twisted C$^{*}$-periodic boundary conditions and show that
they can be used to calculate the monopole mass.

\section{Magnetic charges in the continuum}

\label{sec:continuum}

The most general renormalisable Lagrangian for the SU($N$) gauge
field theory $A_{\mu}$ with an adjoint Higgs field $\Phi$ is \begin{eqnarray}
{\cal L} & = & -{\rm Tr}G^{\mu\nu}G_{\mu\nu}+{\rm Tr}[D_{\mu},\Phi][D^{\mu},\Phi]\nonumber \\
 &  & -m^{2}{\rm Tr}\Phi^{2}-\kappa{\rm Tr}\Phi^{3}-\lambda_{1}({\rm Tr}\Phi^{2})^{2}-\lambda_{2}{\rm Tr}\Phi^{4},\end{eqnarray}
 where we have used the covariant derivative and field strength tensor
defined by \begin{equation}
D_{\mu}=\partial_{\mu}+igA_{\mu},\quad G_{\mu\nu}=-\frac{i}{g}[D_{\mu},D_{\nu}],\end{equation}
 respectively. Both $\Phi$ and $A_{\mu}$ are Hermitian and traceless
$N\times N$ matrices, which can be expanded in terms of the group
generators $T^{A}$,%
\footnote{We use lower case Latin letters for $a=1,\ldots,N$ and upper case
Latin letters for $A=1,\ldots,(N^{2}-1)$. Greek letters represent
Lorentz indices.%
} \begin{equation}
\Phi(x)=\phi^{A}(x)T^{A},\quad A_{\mu}(x)=A_{\mu}^{A}(x)T^{A},\label{equ:Texpand}\end{equation}
 with real coefficients $\phi^{A}$ and $A_{\mu}^{A}$. The fields
can therefore also be thought of as $N^{2}-1$ component vectors.

Let us first consider the case $N=2$. In this case the group generators
can be chosen to be the Pauli matrices, \begin{equation}
T^{A}=\frac{\sigma^{A}}{2}.\end{equation}
 Because of the properties of the Pauli matrices, ${\rm Tr}\Phi={\rm Tr}\Phi^{3}=0$
and $({\rm Tr}\Phi^{2})^{2}=2{\rm Tr}\Phi^{4}$, and therefore we
can choose $\kappa=\lambda_{2}=0$ without any loss of generality.

In the broken phase, where $m^{2}<0$, the Higgs field has a vacuum
expectation value \begin{equation}
\langle{\rm Tr}\Phi^{2}\rangle=\frac{1}{2}\langle\phi^{A}\phi^{A}\rangle=\frac{v^{2}}{2}\equiv\frac{m^{2}}{\lambda}.\end{equation}
 The SU($2$) symmetry is spontaneously broken to U($1$). To represent
the direction of symmetry breaking, we define 
\begin{equation}
\hat{\Phi}(x)=\frac{\Phi(x)}{\sqrt{2{\rm Tr}\Phi(x)^{2}}},\end{equation}
 which is well defined whenever $\Phi\ne0$. Following 't Hooft \citet{'tHooft:1974qc},
we use this to define the field strength 
\begin{eqnarray}
\label{equ:Fmunu}
F_{\mu\nu} & = & 2{\rm Tr}\hat{\Phi}G_{\mu\nu}-\frac{4i}{g}{\rm Tr}\hat{\Phi}[D_{\mu},\hat{\Phi}][D_{\nu},\hat{\Phi}]\nonumber \\
 & = & \partial_{\mu}(\hat{\phi}^{A}A_{\nu}^{A})-\partial_{\nu}(\hat{\phi}^{A}A_{\mu}^{A})+\frac{\epsilon_{ABC}}{g}\hat{\phi}^{A}(\partial_{\mu}\hat{\phi}^{B})\partial_{\mu}\hat{\phi}^{C}.
 \nonumber\\&&
 \end{eqnarray}

Fixing the unitary gauge, in which $\Phi\propto\sigma^{3}$, makes
this definition more transparent. The gauge fixing is achieved by
the gauge transform $R(x)$, such that the transformed field $\tilde{\Phi}$
is diagonal \begin{equation}
\tilde{\Phi}(x)\equiv R^{\dagger}(x)\Phi(x)R(x)=\sqrt{2{\rm Tr}\Phi^{2}}\frac{\sigma^{3}}{2}.\label{equ:SU2diag}\end{equation}
 In terms of the transformed gauge field \begin{equation}
\tilde{A}_{\mu}=R^{\dagger}A_{\mu}R-\frac{i}{g}R^{\dagger}\partial_{\mu}R,\label{equ:Atfm}\end{equation}
 the field strength tensor has the usual Abelian form, \begin{equation}
F_{\mu\nu}=\partial_{\mu}\tilde{A}_{\nu}^{3}-\partial_{\nu}\tilde{A}_{\mu}^{3}.\label{eq:unitaryFuv}\end{equation}
 Alternatively, we can express this in terms of the diagonal elements
of the transformed gauge field, \begin{equation}
F_{\mu\nu}^{a}=\partial_{\mu}\tilde{A}_{\nu}^{aa}-\partial_{\nu}\tilde{A}_{\mu}^{aa}.\label{equ:diagF}\end{equation}
 This defines a two-component vector of field strength tensors, but
tracelessness of $A_{\mu}$ implies $F_{\mu\nu}^{2}=-F_{\mu\nu}^{1}$.
The conventional field strength $F_{\mu\nu}$ is given by $F_{\mu\nu}=F_{\mu\nu}^{1}-F_{\mu\nu}^{2}$.

The conserved magnetic current corresponding to the residual U(1)
group is defined as \begin{equation}
j_{\mu}^{a}=\partial^{\nu}{}^{\star}F_{\mu\nu}^{a},\label{equ:magcurrent}\end{equation}
 where $^{\star}F_{\mu\nu}^{a}$ is the dual tensor, \begin{equation}
^{\star}F_{\mu\nu}^{a}=\frac{1}{2}\epsilon_{\mu\nu\rho\sigma}F^{a\,\rho\sigma}.\label{eq:dualFmunu}\end{equation}
 Like the field strength, the magnetic currents satisfy $j_{\mu\nu}^{2}=-j_{\mu\nu}^{1}$,
so there is only one monopole species.

Substituting Eq.~(\ref{equ:Fmunu}), one finds \begin{equation}
j_{\mu}^{1}=\frac{1}{4g}\epsilon_{\mu\nu\rho\sigma}\epsilon_{ABC}(\partial^{\nu}\hat{\phi}^{A})(\partial^{\rho}\hat{\phi}^{B})(\partial^{\sigma}\hat{\phi}^{C})=-j_{\mu}^{2}.\end{equation}
 This clearly vanishes when $\Phi\ne0$, but is generally non-zero
when $\Phi$ vanishes. The magnetic charge inside volume $V$ bounded
by a closed surface $S$ that encloses a zero is 
\begin{equation}
Q=\int_{V}d^{3}xj_{0}=\pm\frac{2\pi}{g}(1,-1).\end{equation}

This can be generalized to SU($N$) \citet{'tHooft:1981ht}. The matrices
$\{T^{A}\}$ in Eq.~(\ref{equ:Texpand}) are now the generators of
SU($N$) in the fundamental representation, and we assume the usual
normalisation \begin{equation}
{\rm Tr}T^{A}T^{B}=\frac{1}{2}\delta^{AB}.\end{equation}
 As in Eq.~(\ref{equ:SU2diag}), consider a gauge transformation
$R(x)$ that diagonalizes $\Phi(x)$ and places the eigenvalues in
descending order, \begin{equation}
\tilde{\Phi}(x)=R^{\dagger}(x)\Phi(x)R(x)=\text{diag}(\lambda_{1},...,\lambda_{N}),\label{equ:SUNdiag}\end{equation}
 where $\lambda_{1}\ge\lambda_{2}\ge...\ge\lambda_{N}$.

In classical field theory one usually finds that there are only two
distinct eigenvalues, and consequently only one residual U(1) group.
In that case one can use Eq.~(\ref{equ:Fmunu}) to define the corresponding
field strength. However, as we will discuss in Section~\ref{sec:lattice},
in lattice Monte Carlo simulations all the eigenvalues are distinct.
In that case, $\tilde{\Phi}(x)$ is invariant under gauge transformations
generated by the $N-1$ diagonal generators of SU($N$). Thus we are
left with a residual U(1)$^{N-1}$ gauge invariance corresponding
the Cartan subgroup of SU($N$). It is then convenient to follow
't~Hooft~\citet{'tHooft:1981ht} and define the residual U(1) field
strengths by Eq.~(\ref{equ:diagF}), with $a\in\left\{ 1,\ldots,N\right\} $.
The corresponding magnetic currents $j_{\mu}^{a}$ are then given
by Eq.~(\ref{equ:magcurrent}). They satisfy the tracelessness condition
\begin{equation}
\sum_{a=1}^{N}F_{\mu\nu}^{a}=\sum_{a=1}^{N}j_{\mu\nu}^{a}=0,\end{equation}
 so that there are only $N-1$ independent U(1) fields and magnetic
charges.

In three dimensions any two eigenvalues coincide, $\lambda_{b}=\lambda_{b+1}$,
in a discrete set of points, which behave like magnetic
charges with respect to the components $F^b_{\mu\nu}$ and $F^{b+1}_{\mu\nu}$ of the field strength
tensor (\ref{equ:diagF})~\citet{'tHooft:1981ht}.
That is, it behaves like a magnetic monopole with charge $Q=\pm\hat{q}_{b}$,
where the elementary magnetic charges are 
\begin{equation}
\hat{q}_{b}^{a}=\frac{2\pi}{g}\left(\delta_{a,b}-\delta_{a,(b+1)}\right),\label{eq:qa}
\end{equation}
 or in vector notation \begin{equation}
\hat{q}_{b}=\frac{2\pi}{g}\left(\overbrace{0,...,0}^{b-1},1,-1,\overbrace{0,...,0}^{N-b-1}\right).\label{equ:qjdef}\end{equation}
In the core of the monopole, the SU($2$) subgroup involving the $b$th
and $(b+1)$th components of the fundamental representation is
restored.

\section{Magnetic charge on the lattice}

\label{sec:lattice} On the lattice, the Higgs field is defined on
sites $x$ while the gauge degrees of freedom are encoded in SU($N$)
valued link variables $U_{\mu}(x)$. The Lagrangian is given by 
\begin{eqnarray}
\label{equ:latticeL}
{\cal L} & = & \frac{1}{g^{2}}\sum_{\mu\nu}{\rm Tr}U_{\mu}({x})U_{\nu}(x+\hat{\mu})U_{\mu}^{\dagger}({x}+\hat{\nu})U_{\nu}^{\dagger}({x})\nonumber \\
 &  & +2\sum_{\mu}\left[{\rm Tr}\Phi({x})^{2}-{\rm Tr}\Phi({x})U_{\mu}({x})\Phi({x}+\hat{\mu})U_{\mu}^{\dagger}({x})\right]\nonumber \\
 &  & +m^{2}{\rm Tr}\Phi^{2}+\kappa{\rm Tr}\Phi^{3}+\lambda_{1}({\rm Tr}\Phi^{2})^{2}+\lambda_{2}{\rm Tr}\Phi^{4}.\end{eqnarray}

Again, we diagonalise $\Phi$ by a gauge transformation $R(x)$, \begin{equation}
\tilde{\Phi}(x)=R^{\dagger}(x)\Phi(x)R(x).\label{equ:diagPhi}\end{equation}
 Link variables are transformed to \begin{equation}
\tilde{U}_{\mu}(x)=R^{\dagger}(x)U_{\mu}(x)R(x+\hat{\mu}).\end{equation}
 The diagonalised field $\tilde{\Phi}$ is still invariant under diagonal
gauge transformations, 
\begin{eqnarray}
D(x)&=&{\rm diag}\,(e^{i\Delta_{1}(x)},\ldots,e^{i\Delta_{N}(x)}),\nonumber\\
&& \sum_{a=1}^{N}\Delta_{a}=0 \mod~2\pi,
\end{eqnarray}
 which form the residual U(1)$^{N-1}$ symmetry group and contain the
 elements of the center $\mathbb{Z}_N$ of SU($N$).
 
To identify the corresponding U(1) field strength tensors, we need
to decompose $\tilde{U}_{\mu}$~\citet{Kronfeld:1987vd}, 
\begin{equation}
\tilde{U}_{\mu}(x)=C_{\mu}(x)u_{\mu}(x),\label{eq:generaldecomp}\end{equation}
 where $u_{\mu}(x)$ represents the residual U(1) gauge fields and
transforms as \begin{equation}
u_{\mu}(x)\rightarrow D^{\dagger}(x)u_{\mu}(x)D(x+\hat{\mu}),\end{equation}
and $C_{\mu}(x)$ represents fields charged under the U(1) groups.
This decomposition is not unique~\citet{Kronfeld:1987vd}.
A simple choice is to define Abelian link variables as the diagonal
elements of $\tilde{U}_{\mu}$ in direct analogy with Eq.~(\ref{equ:diagF}),
\begin{equation}
u_{\mu}(x)={\rm diag}\,\tilde{U}_{\mu}(x).
\label{eq:specdec}
\end{equation}
 In practice, it is often more convenient to work with link angles
and define an $N$-component vector \begin{equation}
\alpha_{\mu}^{a}(x)=\arg u_{\mu}^{aa}.\label{eq:kronalpha}\end{equation}
 As angles, these are only defined modulo $2\pi$, and we choose them
to be in the range $-\pi<\alpha_{\mu}^{a}\le\pi$. As in the continuum,
the angles $\alpha_{\mu}^{a}$ satisfiy \begin{equation}
\sum_{a}\alpha_{\mu}^{a}(x)=0~\mod~2\pi.\end{equation}
 Therefore it has only $N-1$ independent components, corresponding
to the $N-1$ residual U(1) gauge groups.

Next, we construct plaquette angles as \begin{equation}
\alpha_{\mu\nu}^{a}(x)=\alpha_{\mu}^{a}(x)+\alpha_{\nu}^{a}(x+\hat{\mu})-\alpha_{\mu}^{a}(x+\hat{\nu})-\alpha_{\nu}^{a}(x),\label{eq:plaqangles}\end{equation}
 which are the lattice analogs of the Abelian field strength. In the
continuum limit, they are related by \begin{equation}
F_{\mu\nu}^{a}=\frac{1}{g}\alpha_{\mu\nu}.\end{equation}
 Because the links $\alpha_{\mu}^{a}$ are only defined modulo $2\pi$,
the same applies to the plaquette, and again, we choose $-\pi<\alpha_{\mu\nu}^{a}\le\pi$.

Using Eq.~(\ref{eq:plaqangles}), the corresponding lattice magnetic
currents are 
\begin{equation}
j_{\mu}^{a}=\frac{1}{g}\Delta_{\nu}^f \ ^\star\alpha_{\mu\nu}^{a},\end{equation}
 where $\Delta_{\nu}^f $ is the forward derivative in direction $\nu$
 on the lattice and
\begin{equation}
\ ^\star \alpha_{\mu\nu}^{a}=\frac{1}{2}\epsilon_{\mu\nu\rho\sigma}\alpha_{\rho\sigma}^{a}.\end{equation}
 These are integer multiples of $2\pi$, because each contribution
of $\alpha_{\mu}^{a}(x)$ is cancelled by a $-\alpha_{\mu}^{a}(x)$
modulo $2\pi$.

In particular, the Abelian magnetic charge inside a single lattice
cell is given by \begin{equation}
q^{a}(x)=j_{0}^{a}=\frac{1}{2g}\sum_{ijk}\epsilon_{ijk}\left(\alpha_{ij}^{a}(x+\hat{k})-\alpha_{ij}^{a}(x)\right).\label{equ:qdef}\end{equation}
 Each component of this vector is an integer multiple of $(2\pi/g)$,
and they all add up to zero. The elementary charges, corresponding
to individual monopoles, are the same as in the continuum~(\ref{equ:qjdef}).
Other values of the charge vector $q$ correspond to composite states
made of elementary monopoles.

The diagonalisation procedure in Eq.~(\ref{equ:SUNdiag}) is ill
defined whenever the Higgs field has degenerate eigenvalues, but on
lattice the set of field configurations in which that happens has
zero measure in the path integral. Physically this means that the
core of the monopole never lies exactly at a lattice site. Therefore
these configurations do not contribute to any physical observable and
do not have to be considered separately.

\section{Monopole mass}

\label{sec:mass} The Abelian magnetic charge $Q$ of any lattice
field configuration is well defined by adding up the contributions
(\ref{equ:qdef}) from each lattice cell, \begin{equation}
Q=\sum_{x}q(x).\end{equation}
 Because it is discrete, one can define separate partition functions
$Z_{Q}$ for each magnetic charge sector. The full partition function
is simply the product\[
Z=\prod_{Q}Z_{Q}.\]
 The ground state energy of a given charge sector may be defined by
\begin{equation}
E_{Q}=-\frac{1}{T}\ln\frac{Z_{Q}}{Z_{0}},\end{equation}
 where $Z_{0}$ is the partition function of the charge zero sector
and $T$ is the length of the lattice in the time direction. 
The mass $M_{j}$ of a single monopole $\hat{q}_{j}$
is given by the ground state energy of the corresponding charge sector
\begin{equation}
M_{j}=E_{\hat{q}_{j}}.\end{equation}

In order to calculate the energies $E_{Q}$, we need to impose boundary
conditions that enforce non-trivial Abelian magnetic charge. It is
important that these boundary conditions preserve the translational
invariance of the system, because otherwise our calculations are tainted
by boundary effects. Because they are generally proportional to the
surface area they would completely swamp the contribution from a point-like
monopole which we want to measure.

Gauss's law rules out periodic boundary conditions since they fix
the charge to zero. However, translational invariance only requires
periodicity up to the symmetry of the Lagrangian (\ref{equ:latticeL}).
Since the magnetic current is conserved, we need only consider spatial
boundary conditions.

For SU($2$), it was found in \citet{Davis:2000kv} that the following
boundary conditions force an odd value for the magnetic charge,
\begin{eqnarray}
\Phi(x+L\hat{\jmath}) & = &
-\sigma_{j}\Phi(x)\sigma_{j}=(\sigma_{2}\sigma_{j})^{\dagger}\Phi^{*}(x)
      (\sigma_{2}\sigma_{j}),\nonumber\\  
U_{\mu}(x+L\hat{\jmath}) & = &
\sigma_{j}U_{\mu}(x)\sigma_{j}=
(\sigma_{2}\sigma_{j})^{\dagger}U_{\mu}^{*}(x)(\sigma_{2}\sigma_{j}).
\label{eq:su2bc}\nonumber
\end{eqnarray}
 These are an example of twisted C-periodic boundary conditions, as
introduced by Kronfeld and Wiese \citet{Kronfeld:1990qu}. Note that
while twisted C-periodic and twisted periodic boundary conditions are equivalent
for the gauge links, the Higgs field requires an
additional anti-periodicity when we convert from one form to the other.
Physically, this means that charge conjugation is carried only
by the Higgs field in SU($2$). We will come back to this important point
when we discuss the boundary conditions in terms of the flux sectors
of pure SU($2$) gauge theory.

In contrast, 
it turns out that untwisted C-periodic boundary conditions,
\begin{eqnarray}
\Phi(x+L\hat{\jmath}) & = &
          -\sigma_2\Phi(x)\sigma_2=\Phi^{*}(x)\nonumber \\ 
U_{\mu}(x+L\hat{\jmath}) & = & 
\sigma_2U_{\mu}(x)\sigma_2
=U_{\mu}^{*}(x),\label{eq:su2untwistedbc}
\end{eqnarray}
are compatible with any even value of magnetic
charge~\citet{Davis:2000kv} but are  
locally gauge equivalent to the twisted ones (\ref{eq:su2bc}).
Assuming that monopoles do not form bound states, the weight of the
multi-monopole configurations in the path integral is exponentially
suppressed \begin{equation}
Z_{Q}=e^{-MT}Z_{0},\end{equation}
 where $M$ is the monopole mass and $T$ is the temporal size of
the lattice. In the infinite-volume limit, $T\rightarrow\infty$,
only the configurations with the minimum number of monopoles contribute
to the path integral. So the partition function $Z_{{\rm odd}}$ for
twisted C-periodic boundary conditions will be dominated by configurations
with a single monopole, while the partition function $Z_{{\rm even}}$
will be dominated by configurations with no monopoles. Therefore the
monopole mass is given by 
\begin{equation}
M=-\lim_{T\rightarrow\infty}\frac{1}{T}\ln Z_{{\rm odd}}/Z_{{\rm
    even}}.
\label{eq:Zratio}\end{equation}
This was used to calculate the non-perturbative mass of the 't Hooft-Polyakov
monopole in \citet{Davis:2001mg,Rajantie:2005hi}, with good agreement
with classical expectations.

\section{Twisted boundary conditions}

\label{sec:twist} Let us now generalise the boundary conditions
(\ref{eq:su2bc}) to SU($N$) with $N>2$. To avoid boundary effects,
the boundary conditions must preserve translation invariance, and
they will therefore have to be periodic up to the symmetries of the
theory. In the case of Eq.~(\ref{equ:latticeL}), the available symmetries
are complex conjugation of the fields and gauge invariance. When $\kappa=0$,
reflection of the Higgs field $\Phi\rightarrow-\Phi$ is also a symmetry,
but in general it is not, and therefore we do not consider it. The
appropriate extension of (\ref{eq:su2bc}) is then a combination of
complex conjugation and gauge transformations.

\subsection{Fully C-periodic boundary conditions}

It is natural to impose complex conjugation in all three spatial directions,
in which case we have 
\begin{eqnarray}
\Phi(x+L\hat{\jmath}) & = &
\Omega_{j}^{\dagger}(x)\Phi^{*}(x)\Omega_{j}(x),\nonumber \\ 
U_{\mu}(x+L\hat{\jmath}) & = &
   \Omega_{j}^{\dagger}(x)U_{\mu}^{*}(x)\Omega_{j}(x+\hat{\mu}),
\label{eq:Cbc}\end{eqnarray}
 where the SU($N$) gauge transformation matrix $\Omega_{j}(x)$ can
in general be position dependent. We refer to these as (fully) C-periodic
boundary conditions \citet{Kronfeld:1990qu}.\footnote{In fact, in the
  terminology of Ref.~\citet{Kronfeld:1990qu}, these  
correspond to $C$-periodic boundary conditions with $C=-1$, and
$C=1$ would correspond to boundary conditions without complex
conjugation.} 

To avoid contradiction at the edges, it should not matter in which
order the boundary conditions are applied. Therefore, the gauge transformations
must satisfy~\citet{Kronfeld:1990qu} 
\begin{equation}
\begin{split}
\Omega_{j}^{\dagger} & (x+L\hat{k})\Omega_{k}^{T}(x) 
      \Phi(x)\Omega_{k}^{*}(x)\Omega_{j}(x+L\hat{k})\\
 & =\Phi(x+L\hat{\jmath}+L\hat{k})\\
 & =\Omega_{k}^{\dagger}(x+L\hat{\jmath})\Omega_{j}^{T}(x)
       \Phi(x)\Omega_{j}^{*}(x)\Omega_{k}(x+L\hat{\jmath}),\end{split}
\label{eq:cornerPhi}\end{equation}
 and \begin{equation}
\begin{split}\Omega_{j}^{\dagger} & (x+L\hat{k})\Omega_{k}^{T}(x)U_{\mu}(x)\Omega_{k}^{*}(x+\hat{\mu})\Omega_{j}(x+L\hat{k}+\hat{\mu})\\
 & =U_{\mu}(x+L\hat{\jmath}+L\hat{k})\\
 & =\Omega_{k}^{\dagger}(x+L\hat{\jmath})\Omega_{j}^{T}(x)U_{\mu}(x)\Omega_{j}^{*}(x+\hat{\mu})\Omega_{k}(x+L\hat{\jmath}+\hat{\mu}).\end{split}
\label{eq:cornerU}
\end{equation}
 Since our fields are blind to center elements, Eq.~(\ref{eq:cornerPhi})
implies the cocycle condition 
\begin{eqnarray}
\Omega_{i}^{*}(x) \Omega_{j}(x+ L \hat \imath )
&=& z_{ij} \, \Omega_{j}^{*}(x) \Omega_{i}(x+L\hat\jmath ) , \nonumber\\
 z_{ij}&=&  e^{i \theta_{ij}} ,  
\label{eq:cocycle}
\end{eqnarray}
where the $N$th roots of unity $z_{ij} =z_{ji}^{*}$ are formed by the 
antisymmetric 'twist tensor' $\theta_{ij} = - \theta_{ji}$ with the
usual parametrisation in terms of three $\mathbb Z_N$-valued numbers $m_i$, 
\begin{equation}
\theta_{ij}=\frac{2\pi}{N} \, \epsilon_{ijk} m_k ,\quad
m_{i}\in\mathbb Z_N.\label{equ:ZN}
\end{equation} 
 Furthermore, Eq.~(\ref{eq:cornerU}) implies that the $z_{ij}$ have
to be independent of position. 

All choices of  $\Omega_{i}(x)$, $\Omega_{j}(x)$ with the same twist $z_{ij}$
are gauge equivalent~\citet{'tHooft:1979uj,Kronfeld:1990qu}, and
we therefore assume that we can choose the matrices $\Omega_{j}$ to be 
independent of position analogous to the standard `twist eaters' in the case
of 't Hooft's twisted boundary conditions without charge conjugation
\cite{Ambjorn:1980sm}. Explicit realisations for the allowed
C-periodic twists \cite{Kronfeld:1990qu} by constant $\Omega$'s for
even $N$ are straightforward and will be given in
Sec.~\ref{sec:allowed} below.

The fact that non-trivial twists with C-periodic boundary conditions
are only possible for even $N$ can be seen explicitly by
considering the effect of the cocycle condition (\ref{eq:cocycle})
on the product $\Omega_{i}\Omega_{j}^{*}\Omega_{k}$~\citet{Kronfeld:1990qu}.
On one hand, we have \begin{eqnarray}
\Omega_{i}\Omega_{j}^{*}\Omega_{k} & = & z_{jk}\Omega_{i}\Omega_{k}^{*}\Omega_{j}\nonumber \\
 & = & z_{jk}z_{ki}\Omega_{k}\Omega_{i}^{*}\Omega_{j}\nonumber \\
 & = & z_{jk}z_{ki}z_{ij}\Omega_{k}\Omega_{j}^{*}\Omega_{i},\label{eq:permute1}\end{eqnarray}
 but applying the condition in the opposite order we find \begin{eqnarray}
\Omega_{i}\Omega_{j}^{*}\Omega_{k} & = & z_{ji}\Omega_{j}\Omega_{i}^{*}\Omega_{k}\nonumber \\
 & = & z_{ji}z_{ik}\Omega_{j}\Omega_{k}^{*}\Omega_{i}\nonumber \\
 & = & z_{ji}z_{ik}z_{kj}\Omega_{k}\Omega_{j}^{*}\Omega_{i}.\label{eq:permute2}\end{eqnarray}
 Therefore the twist tensors must satisfy the constraint \begin{equation}
z_{ji}^{2}z_{jk}^{2}z_{ki}^{2}=1,\end{equation}
 which implies for the $\mathbb Z_N$ valued $m_i$, that
\begin{equation}
\smallfrac{2}{N} (m_{1}+m_{2}+ m_{3})\in \{0,1\} .
\label{eq:z constraint}
\end{equation} 
Hence, for non-trivial C-periodic twist, 
$N/2$ must be in $ \mathbb{Z}_N$, i.e. $N$ must be even
\cite{Kronfeld:1990qu}.  

Let us now consider the effect of the boundary conditions (\ref{eq:Cbc})
on the residual U(1) fields. Because the eigenvalues of the Higgs
field $\Phi$ don't change under the twists in (\ref{eq:Cbc}), i.e. $\Phi(x)$
and $\Phi(x + L \hat\jmath )$ have the same set of eigenvalues, which
are all real, we can choose  the diagonalised field $\tilde{\Phi}$  
defined in Eq.~(\ref{equ:diagPhi}) to be periodic,
\begin{equation}
   \tilde\Phi(x + L \hat\jmath ) = \tilde\Phi(x) .
\end{equation}
Then, on one hand, 
\begin{eqnarray}
\Phi({x}+L\hat{\jmath}) & = & \Omega_{j}^{\dagger}\Phi^{*}({x})\Omega_{j}\nonumber \\
 & = & \Omega_{j}^{\dagger}\left(R({x})\tilde{\Phi}({x})R^{\dagger}({x})\right)^{*}\Omega_{j}\nonumber \\
 & = &
 \Omega_{j}^{\dagger}R^{*}({x})\tilde{\Phi}({x}+L\hat{\jmath})R^{T}({x})\Omega_{j},  
\end{eqnarray}
while on the other,
\begin{equation}
\Phi({x}+L\hat{\jmath})=R({x}+L\hat{\jmath})
\tilde{\Phi}({x}+L\hat{\jmath})R^{\dagger}({x}+L\hat{\jmath}) .
\end{equation} 
To ensure the compatibility of the two, we impose spatial boundary
conditions for $R(x)$ as follows,  
%Compatibility of the two implies spatial boundary conditions for
%$R(x)$ as follows, 
\begin{equation} 
R({x}+L\hat{\jmath})=\Omega_{j}^{\dagger}R^{*}({x}).
\end{equation}
 When we apply multiple translations by $L$, however, we observe $\mathbb{Z}_N$
 jumps in the definition of gauge transforms $R(x)$ in SU($N$). A
 double translation by $L$ first along the $k$ direction and then
 along $j$ for example is defined by
\begin{equation} 
R^{jk}({x}+L\hat{\jmath}+L\hat{k}) \equiv
\Omega_{j}^{\dagger}\Omega_{k}^{T}R({x}) ,
\end{equation}
while for a translation first along $j$ followed by one in the $k$
direction leads to
\begin{equation} 
R^{kj}({x}+L\hat{\jmath}+L\hat{k}) \equiv
\Omega_{k}^{\dagger}\Omega_{j}^{T}R({x}) ,
\end{equation}
From (\ref{eq:cocycle}) it then immediately follows that
\begin{equation}
R^{jk}({x}+L\hat{\jmath}+L\hat{k}) = z_{kj}
R^{kj}({x}+L\hat{\jmath}+L\hat{k}) .  \label{eq:Rtwist}
\end{equation}
From their effect in (\ref{equ:diagPhi}), or generally in
SU($N$)$/\mathbb{Z}_N$,
% without fundamental or other fields that
%represent the center non-trivially, 
these two would be equivalent. In  SU($N$) they are not, however. 
There, transformations where the $R's$ applied at a corner site to
links in different directions attached to that corner differ, by center
elements as in (\ref{eq:Rtwist}), can be used to change the twist
sector. If we allowed such multi-valued, and hence singular gauge
transformations, we could then arrange matters such that the
transformed link variables $\tilde U$ would all be C-periodic,  
\begin{eqnarray}
\tilde{U}_{\mu}({x}+L\hat{\jmath}) & = &
R^{\dagger}({x}+L\hat{\jmath})U_{\mu}({x}+L\hat{\jmath})
R({x}+\hat\mu+L\hat{\jmath})\nonumber   \\ 
 & = & R^{T}({x})U_{\mu}^{*}({x})R^{*}({x}+\hat{\mu})\nonumber \\
 & = & \tilde{U}_{\mu}^{*}({x}) . \label{eq:cptildeU}
\end{eqnarray}
The twist would then be completely removed by the singular gauge
transformation, however. Conversely, when comparing a fundamental
Wilson loop that winds around a plane with non-trivial twist to the 
corresponding loop formed by the $\tilde U$'s, one would observe that the
original loop obtained its center flux entirely from the $\mathbb{Z}_N$
jump of the multi-valued gauge transformation, while the $\tilde U$ loop,
with purely C-periodic b.c.'s (\ref{eq:cptildeU}), would be trivial.   

In order to preserve the  $\mathbb{Z}_N$ center flux in SU($N$), we
must apply single-valued and hence proper SU($N$) gauge
transformations, without such a jump. Those will of course not change
the Wilson loop at all, when transforming the $U$'s to the gauge-fixed
links $\tilde U$.    
Then however, we have to decide how we define the gauge transformation
at those corner sites where $\mathbb{Z}_N$ ambiguities as in (\ref{eq:Rtwist})
arise. Consequently, the boundary conditions (\ref{eq:cptildeU}) for
the gauge-fixed $\tilde U$'s attached to such a corner will
have to be amended. 
 
This is best exemplified in two dimensions (with two integer coordinates $x$
and $y$ both ranging from $0$ to $L-1$): At the site with
coordinates $(L,L)$ we define the gauge transformation $R$ as, say
\begin{equation}
   R(L,L) \equiv \Omega_y^\dagger \Omega_x^T R(0,0) = \Omega_y^\dagger
   R^*(L,0) .
\end{equation}
If we consider the $x$ link attached to this corner site, we obtain the
boundary condition
\begin{eqnarray}
  \label{eq:2dconrer1}
  \tilde U_x(L-1,L) &=& R^\dagger(L-1,L) U_x(L-1,L) R(L,L) \nonumber\\
                    &=& R^T(L-1,0) U_x^*(L-1,0) R^*(L,0)
                    \nonumber\\
                    &=& \tilde U_x^*(L-1,0) ,
\end{eqnarray}
as in (\ref{eq:cptildeU}) and as for every other link that is not
connected to this corner. For the corresponding $y$ link at this
corner on the other hand,
\begin{eqnarray}
  \label{eq:2dconrer2}
  \tilde U_y(L,L-1) &=& R^\dagger(L,L-1) U_y(L,L-1) R(L,L) \nonumber\\
      && \hskip -2cm  = R^T(0,L-1) 
                  U_y^*(0,L-1) \Omega_x \Omega_y^\dagger \Omega_x^T
                  R^*(0,0) \nonumber\\
      && \hskip -2cm  = z_{21} \, R^T(0,L-1) U_y^*(0,L-1) R^*(0,L)
                    \nonumber\\
      && \hskip -2cm  = z_{21} \, \tilde U_y^*(0,L-1) ,
\end{eqnarray}
because $\Omega_y^\dagger \Omega_x^T = z_{21} \Omega_x^\dagger
\Omega_y^T $ and $R^*(0,L) = \Omega_y^T R(0,0)$.
This shows that all but one of the gauge-fixed links in the plane are C-periodic
(\ref{eq:cptildeU}) and that the center flux comes about by the
boundary condition of the one link remaining.   

\begin{figure}[t]

\includegraphics[height=6cm,clip=true]{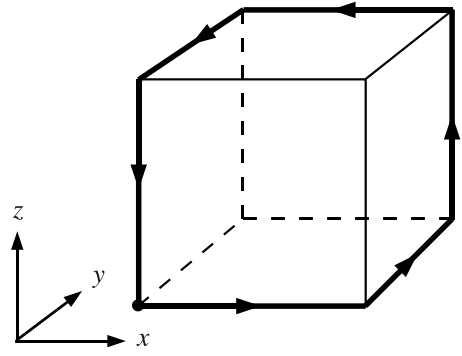}

\caption{Integration curve used to calculate the flux through half of the box.
\label{fig:halfcurve}}

\end{figure}

\begin{figure}[t]

\includegraphics[width=\linewidth,clip=true]{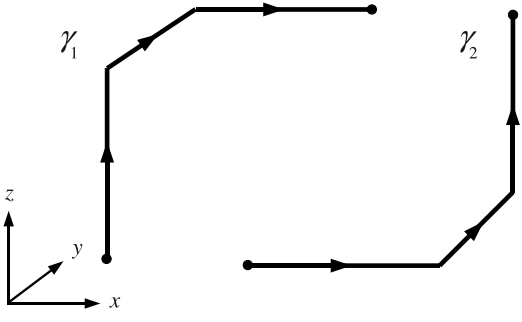}

\caption{Two line segments $\gamma_1$ and $\gamma_2$ that can be used
  to compose the loop in Fig.~\ref{fig:halfcurve}.}
\label{fig:HalfPaths}

\end{figure}

In the following we will only consider proper transformations $R$,
single-valued in SU($N$), so that the center flux is preserved
in the gauge-fixed links, $\tilde U$. In higher dimensions we
therefore introduce the convention that for gauge transformations $R$
involving multiple translations by $L$ these translations are always 
applied in lexicographic order. In three dimensions, this 
leads to the following definitions for the far edges of our $L^3$ box
with one corner in the origin at (0,0,0),
\begin{eqnarray}
    R(L,L,z) &\equiv & \Omega^\dagger_y \Omega^T_x R(0,0,z), \nonumber\\
    R(L,y,L) &\equiv & \Omega^\dagger_z \Omega^T_x R(0,y,0), \nonumber\\
    R(x,L,L) &\equiv & \Omega^\dagger_z \Omega^T_y R(x,0,0), 
\end{eqnarray}
where $x$, $y$ and $z$ run from $0$ to $L-1$; and for the corner
diagonally opposite to the origin, we use
\begin{equation}
  \label{eq:farcornerbc}
      R(L,L,L) \, \equiv \, \Omega^\dagger_z \Omega^T_y \Omega^\dagger_x
    R^*(0,0,0) .
\end{equation}
In particular, we then have
\begin{equation}
  \label{eq:tripLbc}
  R(L,L,L) = z_{12} z_{23} z_ {31} \Omega^\dagger_x \Omega^T_y \Omega^\dagger_z
    R^*(0,0,0)  ,
\end{equation}
and the factor 
\begin{equation}
  \label{eq:totFlux}
  z_{12} z_{23} z_ {31} = \exp\big\{ \smallfrac{2\pi i}{N} (m_1+m_2+m_3)\big\}
\end{equation}
represents the total center flux as measured by a maximal-size Wilson
loop $W(C)$ along the corners of the three-dimensional cube that cuts its
surface into two equal halfs as in Fig.~\ref{fig:halfcurve}. To see
this, let the loop $C$ in Fig.~\ref{fig:halfcurve} be composed of two
line segments $-\gamma_1$ and $\gamma_2$ as shown in
Fig.~\ref{fig:HalfPaths}, for example, and consider gauge transforming
the two Wilson lines $W(\gamma_1)$ and $W(\gamma_2)$. To make them
equal, so that $W(C)= W(\gamma_2)W^\dagger(\gamma_1) = 1$, we would need to
apply a gauge transform 
\[ R^{(1)}(L,L,L) =  \Omega^\dagger_z \Omega^T_y 
\Omega^\dagger_x R^*(0,0,0)\]
at the end of line $W(\gamma_1)$, but
\[ R^{(2)}(L,L,L) =  \Omega^\dagger_x \Omega^T_y \Omega^\dagger_z
R^*(0,0,0) \] 
at the end of $W(\gamma_2)$. This would be a multi-valued
gauge transform with a jump at the far
corner at $(L,L,L)$, however. If we apply the same $R(L,L,L) \equiv
R^{(1)}(L,L,L)$ at the end of both lines, $W(\gamma_1) $ and
$W(\gamma_2)$, the loop $W(C)$ remains unchanged, and we have, 
\begin{equation} 
  W(C) =  W(\gamma_2)W^\dagger(\gamma_1) =   z_{12} z_{23} z_
  {31} \label{eq:WilsFlux}\ . 
\end{equation}

\begin{figure}[t]

\includegraphics[width=\linewidth,clip=true]{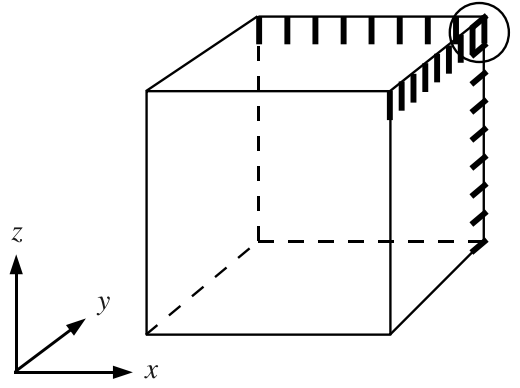}

\caption{Illustration of links with special boundary conditions in
  three dimensions. When center flux is moved to the upper right plaquettes of
  all two dimensional planes, it piles up near the corner at $(L,L,L)$
  as highlighted by the circle.}
\label{fig:TaggedLinks}

\end{figure}

\noindent
In terms of the gauge-fixed links $\tilde U$, we then still have C-periodic
boundary conditions (\ref{eq:cptildeU}) for most of the links,
but we need to take into account the following exceptions:
\begin{eqnarray}
  \tilde U_y(L,L-1,z) &=&  z_{21} \, \tilde U_y^*(0,L-1,z) ,
  \nonumber\\
  \tilde U_z(L,y,L-1) &=&  z_{31} \, \tilde U_z^*(0,y,L-1) ,
  \nonumber\\
  \tilde U_z(x,L,L-1) &=&  z_{32} \, \tilde U_z^*(x,0,L-1)
  , \label{eqs:tUspecbc} 
\end{eqnarray}
where the third variable runs from 0 to $L-1$ again; and, from the
corner at $(L,L,L)$,
\begin{eqnarray}
  \tilde U_y(L,L-1,L) &=&  z_{12} \, \tilde U_y(0,L-1,0) ,
\label{eq:sepcUtbc_2a}\\
  \tilde U_z(L,L,L-1) &=&  z_{32} z_{13} \, \tilde U_z(0,0,L-1) .
  \nonumber 
\end{eqnarray}
The set of special links whose boundary conditions are modified
by center elements is sketched in Fig.~\ref{fig:TaggedLinks}.

Along the loop of Fig.~\ref{fig:halfcurve} this means that almost every link 
in the first half of the loop has a partner in the opposite direction
in the second half, to which it is related by two successive
C-periodic translations (\ref{eq:cptildeU}), and hence periodic. 
There are only two exceptions from the set of twisted links that the
loop picks up. These are 
\begin{eqnarray}
  \label{eq:specUbc_1}
  \tilde U_y(L,L-1,0) &=& z_{21} \tilde U_y^*(0,L-1,0) \\
   &=& z_{21} \tilde U_y(0,L-1,L) , \nonumber
\end{eqnarray}
and the last link of the first half of the loop which ends at the
corner at $(L,L,L)$, as given in (\ref{eq:sepcUtbc_2a}), 
\begin{equation}
  \label{eq:specUbc_2}
  \tilde U_z(L,L,L-1) =  z_{32} z_{13} \tilde U_z (0,0,L-1) .
\end{equation}
The combined center elements are again responsible for the same total
center flux through the loop, now in terms of the gauge-fixed links, 
$\tilde U$.

\subsection{Magnetic flux}

If the decomposition
(\ref{eq:generaldecomp}) commutes with complex conjugation, which
(\ref{eq:specdec}) does, the boundary
conditions (\ref{eq:Cbc}) imply anti-periodicity of the
$\alpha_{\mu}^{a}(x)$ in (\ref{eq:kronalpha}). Therefore the
Abelian projected fields inherit anti-periodic boundary conditions 
\begin{equation}
\alpha_{\mu}^{a}({x}+L\hat{\jmath})=-\alpha_{\mu}^{a}({x}),\label{eq:abelian fields bc}\end{equation}
 except for the special cases, in three dimensions corresponding to the
 links in Eqs.~(\ref{eqs:tUspecbc}), where  
\begin{eqnarray}
 \alpha_y^a(L,L-1,z) &=&  -  \alpha_y^a(0,L-1,z) - \smallfrac{2\pi}{N}\,
 m_3 ,  \label{alphaspecbc_1} \\  
 \alpha_z^a(L,y,L-1) &=&  -  \alpha_z^a(0,y,L-1) + \smallfrac{2\pi}{N}\,
 m_2 , \nonumber\\  
 \alpha_z^a(x,L,L-1) &=&  -  \alpha_z^a(x,0,L-1) - \smallfrac{2\pi}{N}\,
 m_1 , \nonumber
\end{eqnarray}
and in Eqs.~(\ref{eq:sepcUtbc_2a}), where
\begin{eqnarray}
 \alpha_y^a(L,L-1,L) &=& \alpha_y^a(0,L-1,0) + \smallfrac{2\pi}{N}\, m_3 ,
  \label{alphaspecbc_2} \\
 \alpha_z^a(L,L,L-1) &=& \alpha_z^a(0,0,L-1) - \smallfrac{2\pi}{N}\,
 (m_1+m_2)  . \nonumber
\end{eqnarray}
It can be verified that the fluxes in three dimensions, 
\begin{equation}
\alpha_{ij}^{a}(\vec x)=\alpha_{i}^{a}(\vec x)+\alpha_{j}^{a}(\vec
x+\hat{\imath})-\alpha_{i}^{a}(\vec
x+\hat{\jmath})-\alpha_{j}^{a}(\vec x)
\label{equ:plaquettebc}
\end{equation}
 are all essentially anti-periodic, because the twist angles $(2\pi/N)
 m_i$ cancel when we compare fluxes on opposite sides of the lattice. 
There is a single exception here also, however, for which we obtain,
\begin{eqnarray}
  \alpha_{23}(L,L-1,L-1) &=& \\
&& \hskip -3cm  - \alpha_{23} (0,L-1,L-1) -
  \smallfrac{2\pi}{N} \, 2(m_1+m_2+m_3) . \nonumber
\end{eqnarray}
Because of the constraint on the possible twists in Eq.~(\ref{eq:z
  constraint}), and because flux is only defined modulo $2\pi $, 
the additional contribution has no effect, and this is 
equivalent to anti-periodic boundary conditions also. We therefore
have fully anti-periodic Abelian field strengths. This means
that when we cross the boundary we enter a charge conjugated copy
of the same lattice from the opposite side. 

To determine the magnetic charge we repeat the trick of \citet{Davis:2000kv}.
The curve shown in Figure \ref{fig:halfcurve} divides the boundary into
two halves. We denote the magnetic flux through them by $\Phi_{+}$
and $\Phi_{-}$ choosing the positive direction to be pointing outwards.
The two halves are related by the boundary conditions, and in particular,
the antiperiodicity (\ref{equ:plaquettebc}) of the field strength
implies that they are equal $\Phi_{-}=\Phi_{+}$. The magnetic charge
inside the lattice is given by the total flux, which is the sum of
the two contributions, which means \begin{equation}
Q=\Phi_{+}+\Phi_{-}=2\Phi_{+}.\end{equation}
Applying Stokes's theorem, we can write 
\begin{eqnarray}
\Phi_{+}^{a} & = &
  - \frac{1}{g}\left(\sum_{x=0}^{L-1}\alpha_{x}^{a}(x,0,0)
 + \sum_{y=0}^{L-1}\alpha_{y}^{a}(L,y,0) \right.\nonumber \\
 &  & \left.+\sum_{z=0}^{L-1}\alpha_{z}^{a}(L,L,z)
  -\sum_{x=0}^{L-1}\alpha_{1}^{a}(x,L,L)\right.\nonumber \\
 &  &
 \left. -\sum_{y=0}^{L-1}\alpha_{y}^{a}(0,y,L)
 -\sum_{z=0}^{L-1}\alpha_{3}^{a}(0,0,z)\right).
       \label{eq:half flux 1} 
\end{eqnarray}
 When we apply the boundary conditions, all terms cancel except those
involving the cases, 
\begin{eqnarray}
\Phi_{+}^{a} & = & - \frac{1}{g}\left(\alpha_{y}^{a}(L,L-1,0)
  +\alpha_{z}^{a}(L,L,L-1) \right.\nonumber \\
 &  & \hskip .5cm \left. - \alpha_{y}^{a}(0,L-1,L)-\alpha_{z}^{a}(0,0,L-1)
 \right)\nonumber \\ 
 & = & \frac{1}{g} \frac{2\pi}{N} \left( m_{1}+ m_{2} + m_{3}\right) . \label{eq:half flux 2}\end{eqnarray}
where we have used the first equation in (\ref{alphaspecbc_1}) with
$z=0$ and $ \alpha_y^a(0,L-1,0) = - \alpha_y^a(0,L-1,L) $,
and the second equation in (\ref{alphaspecbc_2}).
 
Because the link angles $\alpha_{\mu}^{a}$ are defined modulo
$2\pi$, the fluxes $\Phi_\pm$ are only defined modulo
$(2\pi/g)$. Therefore we find  
\begin{equation}
Q^{a}=\frac{4\pi}{g N}(m_{1}+m_{2}+ m_{3})  \mod~\frac{4\pi}{g} .
\label{equ:Qatheta}
\end{equation}

\subsection{Allowed magnetic charges \label{sub:Allowed-magnetic-charges}}

\label{sec:allowed} It is obvious from Eq.~(\ref{equ:Qatheta})
that the possible charges one can create using the boundary conditions
is quite restricted. As in the continuum (\ref{equ:qjdef}), the components
are quantised in units of $2\pi/g$. 
 Substituting the constraint on the twists for even $N$ in
 Eq.~(\ref{eq:z constraint}) into Eq.~(\ref{equ:Qatheta}) gives the charge
quantisation condition 
\begin{equation}
Q^{a}=\frac{2\pi}{g}\, \mathbb{Z}_2 ,  
\label{equ:Qaquant}\end{equation}
up to integer multiples of $(4\pi/g)$.
Because all components of the charge vector $Q^a $ are furthermore the
same, modulo $(4\pi/g)$, 
we then automatically satisfy the constraint, 
\begin{equation}
   \sum_{a} Q^{a}=NQ_{a}=0\mod~\frac{4\pi}{g}.
\end{equation}
In summary, this means that we can use twised C-periodic boundary
conditions in SU($N$), when $N$ is even, to restrict the ensemble to
either of two distinct classes of monopole configurations. 
If the allowed twists satisfy $m_1+m_2+m_3 = 0$ (modulo $N$), then
their total charges are all integer multiples of $(4\pi/g)$,   
\begin{equation}
Q^{a}=0\mod~\frac{4\pi}{g}~\mbox{for all $a$} . \label{equ:evenQa}
\end{equation}
If the twists are such that  $m_1+m_2+m_3 = N/2$ on the other hand, 
every component of the total charge vector is a half-odd integer
multiple of $(4\pi/g)$,      
\begin{equation}
Q^{a}=\frac{2\pi}{g}\mod~\frac{4\pi}{g}~\mbox{for all $a$}.\label{equ:oddQa}
\end{equation}
Those two sectors differ by at least one unit of Abelian magnetic charge
$(2\pi/g)$ (modulo $(4\pi/g)$) in each of the $N-1$ U($1$)'s. 
This may be due to a single monopole in a diagonally embedded U($1$)
or due to several monopoles in different U($1$)'s depending on the
symmetry breaking pattern. If the symmetry breaking is maximal, these
could be $N-1$ individual monopoles, one in every  U($1$) factor of
the maximal Abelian subgroup of SU($N$).
Because $N$ must be even, the total number of monopoles in the twisted
sector will be odd in either case. The ratio of partition functions of the
two sectors in the infinite volume limit determines the free energy of
such monopole configurations or, at zero temperature, their total mass
as discussed in Section~\ref{sec:mass}.

For even $N$, we can therefore force an odd number of 
monopoles in each residual $U(1)$ by imposing boundary conditions that
correspond to Eq.~(\ref{equ:oddQa}). A convenient choice is 
\begin{equation} 
\begin{array}{c}
\Omega_{1}=\text{diag}(i\sigma_{3},...,i\sigma_{3})\\
\Omega_{2}=\text{diag}(I,...,I)\\
\Omega_{3}=\text{diag}(i\sigma_{1},...,i\sigma_{1}).\end{array}\label{eq:suNbc}\end{equation}
 These are simply the SU(2) matrices from Eq.~(\ref{eq:su2bc}) repeated
in block diagonal form. They satisfy 
\begin{equation}
\Omega_{i}^{*}\Omega_{j}=-\Omega_{j}^{*}\Omega_{i},\; i\ne j,
\end{equation}
 corresponding to a $\pi$ twist angle in each plane, i.e. $m_1=m_2=m_3
 = N/2$. We could equally
well use a a single twisted plane by replacing $\Omega_{1}$ or $\Omega_{3}$
by the unit matrix $1$. An even number of monopoles, corresponding
to Eq.~(\ref{equ:evenQa}), is of course obtained by simply
choosing 
\begin{equation} 
\Omega_{1}=\Omega_{2}=\Omega_{3}=1.\label{equ:notwist}
\end{equation}

We have therefore found that the twisted boundary conditions (\ref{eq:Cbc})
allow us to impose a non-zero magnetic charge, but with several restrictions.
It is, in fact, fairly natural that we cannot specify the exact charge
but only whether it is odd or even with boundary conditions that
preserve translational invariance \cite{Polley:1990tf}.

The other restriction, that all charges must have the same value,
arises because our boundary conditions are linear operations on the
fields. The transformation matrices $\Omega_{j}$ are therefore independent
of the direction of symmetry breaking $\Phi$, which defines the different
residual U(1) groups. Therefore the boundary conditions cannot treat
any U(1) group differently from the others. It may be possible to
avoid this restriction by considering non-linear transformations.
In principle, one could specify the boundary conditions in the unitary
gauge in which the different U(1) groups can be treated separately.
However, it is not clear if it is possible even then to impose translation
invariant boundary conditions that give different values to different
magnetic charges.

In summary, the boundary conditions (\ref{eq:Cbc})
allow us to define the partition functions $Z_{{\rm odd}}$ and $Z_{{\rm even}}$
in Eq.~(\ref{eq:Zratio}) using the gauge transformation (\ref{eq:suNbc})
and (\ref{equ:notwist}), respectively. Using Eq.~(\ref{eq:Zratio}), we
can therefore calculate the energy difference between these two sectors.

If there is only one residual U(1) group, which is usually the case,
only the monopole species that corresponds to it is massive, and therefore
Eq.~(\ref{eq:Zratio}) gives that monopole's mass, just as in SU(2).
If there are several residual U(1) groups, there is a magnetic charge
corresponding to each U(1) group, and therefore $Z_{{\rm odd}}$
generally represents a multi-monopole state. Depending on which
configuration has the lowest energy, the monopoles may either be
separate free particles, in which case Eq.~(\ref{eq:Zratio}) gives the
sum of their masses, or as a bound state in which case it gives the
energy of the bound state.

\subsection{Mixed boundary conditions}
\label{subsec:Mixbc}

In the previous subsection, we 
imposed complex conjugation in all three directions. This has the advantage
of preserving the invariance of the theory under 90-degree rotations.
However, for a non-zero magnetic charge, it is
enough to have complex conjugation in one direction, so that the flux
can escape through at least one face, and we can ask whether that would lead 
to fewer restrictions for the
allowed magnetic charges. In Appendix~\ref{app:mixed} we show that
this is not the case, and that even for such ``mixed'' boundary
conditions, the allowed magnetic charges are constrained exactly as in
Section~\ref{sec:allowed}. 

With a single C-periodic direction, for example, we find in
App.~\ref{app:Cpx} that the outward fluxes, $\Phi^\parallel_\pm$,
parallel to this direction and through the perpendicular faces at
opposite sides of the volume are equal, and quantised in terms of the
magnetic center flux $m_\parallel $ in that direction, 
\begin{equation}
  \Phi_+^{\parallel} =  \Phi_-^{\parallel} = \smallfrac{2\pi}{gN} \,
  m_\parallel \; . \label{eq:Flux1dPar}
\end{equation}
In contrast, the Abelian fluxes in the two orthogonal directions are
no-longer quantised, but they are both conserved, 
\begin{equation}
  \Phi_+^{\perp} +  \Phi_-^{\perp} = 0 \; . \label{eq:Flux1dPerp}
\end{equation}
So there is one extra source of strength $2 m_\parallel/N $ in units
of magnetic charge $(2\pi/g)$ whose entire flux goes along the
C-periodic direction, 
\begin{equation}
  Q  =  \Phi_+^{\parallel} + \Phi_-^{\parallel} = \smallfrac{2\pi}{g} \,
  \smallfrac{2m_\parallel}{N} \; , \label{eq:chargeCpPar}
\end{equation}
again modulo $(4\pi/g) $ and the same for all $a = 1, \dots N-1$.
But as we show in the Appendix, we again have 
\begin{equation} 
  \smallfrac{2m_\parallel }{N} \, \in \{0,1\} ,  
\end{equation}
which again restricts the construction to even $N$, because  $2
m_\parallel/N $ is also an element of the center, $\mathbb Z_N$.

It is instructive to compare this to the case of standard twisted
boundary conditions, without any C-periodic direction, where
the analogously defined Abelian projected fluxes are all quantised
{\em and} conserved, i.e. where
\begin{equation} 
  \Phi^{(i)}_+ = - \Phi_-^{(i)} = \smallfrac{1}{g} \smallfrac{2\pi}{N}
  \, m_i \; , \;\; \mbox{for all} \; i = 1,2,3 . \label{eq:Phitbc}
\end{equation}
The introduction of one C-periodic direction thus led to non-quantised
contributions of Abelian projected flux in the orthogonal directions
in addition to the center flux (\ref{eq:Phitbc}) of the corresponding
sectors with standard twists a la 't Hooft, c.f. Eqs.~(\ref{eq:Phi2})
and (\ref{eq:Phi3}). These non-quantised contributions are
due to the Abelian projection and may not have any physical
significance at all. So unlike standard center flux, the flux in the
orthogonal directions is no-longer quantised, but like standard center
flux it is still conserved. 

In contrast, the flux along the C-periodic direction is still quantised
in units of center elements, see Eqs.~(\ref{eq:Phi1p}) and
(\ref{eq:Phi1m}), but it is no-longer conserved. The introduction of
the C-periodic direction has led to a reversal of the center flux when
passing through the volume along this direction, by introducing a
source of a strength of twice that magnetic flux into the volume,
c.f.~Eq.~(\ref{eq:chargeCpPar}). 

But this only works for center fluxes with $-m = m $, which can be
non-trivial only when $-1$ is among the roots 
of unity and $N$ is even. Then however, these particular fluxes are
$\mathbb Z_2$ valued and do not have a direction. In the pure gauge
theory we cannot even distinguish positive from negative flux in this
case, which is why we can reverse it without harm in the first
place. So for the pure gauge theory we have gained nothing new here.
Moreover, 't Hooft's magnetic fluxes as employed here play no role in the
deconfinement transition of the pure gauge theory, the free energy of
the corresponding center vortices always vanishes in the thermodynamic
limit \cite{vonSmekal:2002gg}.\footnote{Note that combinations of 
  magnetic with electric twists can be used, however, to force
  fractional topological charge and to measure the topological susceptibility
  without cooling \cite{vonSmekal:2002uz}.}

But together with our adjoint Higgs fields, which have 
anti-periodic Abelian components in such a C-periodic direction, we
can distinguish the relative orientations of center vortex and Higgs
field as described in Sec.~\ref{subsec:VortexPicture} below. And
together with the adjoint Higgs field, the different magnetic sectors
have now become relevant -- not for confinement in the pure gauge
theory, but for the masses of 't Hooft-Polyakov monopoles and the
Higgs mechanism.

\section{Relation to vortices}

\label{sec:ReltoVor}

\subsection{The continuum and zeroes of the Higgs field for SU($2$)}

It would be nice to see how the boundary conditions (\ref{eq:Cbc})
relate to magnetic charge in the continuum theory. This is straightforward
when the gauge group is SU($2$). In this case, Abelian monopoles
are located at zeroes of the Higgs field. So to have an odd number
of monopoles we must have an odd number of zeroes of $\Phi$. To proceed,
it's helpful to write the boundary conditions in the form

\begin{equation}
\Phi(x+L\hat{\jmath})=-\sigma_{j}\Phi(x)\sigma_{j}.\end{equation}
It's then clear that the components of the Higgs field in the adjoint
representation $\Phi={\phi}^{A}{\sigma}^{A}/2$ inherit the conditions
\begin{eqnarray}
\phi^{1}(x+L\hat{x}) & = & -\phi^{1}(x),\nonumber \\
\phi^{2}(x+L\hat{y}) & = & -\phi^{2}(x),\nonumber \\
\phi^{3}(x+L\hat{z}) & = & -\phi^{3}(x),\label{eq:all twist cpts}\end{eqnarray}
with all other components periodic. This respects a 'hedgehog' configuration,
as it should.

Note, for example, that $\phi^{1}$ must have an odd number of zeroes
on every line through the box in the $x$ direction. By continuity,
these combine to form surfaces pinned to the boundary of the othogonal
plane. Similarly, there must be an odd number of surfaces through
the $y$ and $z$ directions where $\phi^{2}$ and $\phi^{3}$ are
respectively zero. Because of their relative orthogonality, these
surfaces intersect in an odd number of points where all three components
are zero. To help picture this, consider the surfaces where $\phi^{1}$
and $\phi^{2}$ are zero. These intersect to form an odd number of
lines in the $z$ direction on which $\phi^{1}$ and $\phi^{2}$ are
both zero. Since $\phi^{3}$ is antiperiodic in the $z$ direction, there must be an odd number of points on these lines (and in total) where ${\phi}^{A}$ vanishes. 

All of the (partial or mixed) C-periodic boundary conditions that force
an odd magnetic charge have this property. Conversely, those with
trivial magnetic charge modulo $4\pi$ are found to permit only an
even number points where the Higgs field is zero.

\subsection{Vortex picture - Laplacian center gauge}

\label{subsec:VortexPicture}
As we've seen, the allowed Abelian magnetic charges are tightly connected to and restricted by the center flux sectors of the pure gauge theory. Here the relevant objects are center vortices, which are strings of center flux in three dimensions, and surfaces in four dimensions.

It is commonly believed that colour confinement is the result of certain topological objects that dominate the QCD vacuum on large distance scales, and center vortices are a leading candidate \citet{Greensite:2003bk}.  In the vortex picture of confinement, Wilson loops acquire
a 'disordering' phase factor from every vortex that they link with
\citet{Greensite:2003bk}. The area law for timelike Wilson loops in
pure SU($N$) gauge theory comes from the percolation of spacelike
vortex sheets in the confined phase. Their free energies have been measured
over the deconfinement phase transition in the pure SU(2) gauge theory
with methods entirely analogous to the ones described here, from
ratios of partition functions with twisted boundary conditions in
temporal planes forcing odd numbers of $\mathbb Z_2$ center vortices
through those planes over the periodic ensemble with even numbers
\cite{deForcrand:2001nd,vonSmekal:2002ps}.  
 A Kramers-Wannier duality is then observed by comparing the
behaviour of these center vortices with that of 't Hooft's electric
fluxes which yield the free energies of static charges in
a well-defined (UV-regular) way~\cite{deForcrand:2001dp}, with
boundary conditions to mimic the presence of 'mirror' (anti)charges in
neighbouring volumes. This duality follows that between the Wilson
loops of the 3-dimensional $\mathbb Z_2$-gauge theory and the 3d-Ising spins,
reflecting the universality of the center
symmetry breaking transition. 

This is in contrast to the monopole scenario, where confinement is attributed
to the dual Meissner effect from a monopole condensate.
It turns out that these descriptions may be complimentary, at least in
certain gauges. In the last
few years it's become clear that monopole world lines are embedded
on the surface of center
 vortices \citet{Ambjorn:1999ym,Chernodub:2004em,deForcrand:2000pg,Reinhardt:2001kf,%
Cornwall:1999xw,Cornwall:1998ef,Cornwall:2001ni}.
Percolation of one implies percolation of the other. From this perspective,
we can regard center vortices as Abelian vortices, sourced by the
monopoles. For SU($2$) gauge theory, the monopoles are like beads
on a necklace. For general SU($N$), several center vortices may meet
at a point and we instead have monopole-vortex nets. Similar objects
have been found in various supersymmetric gauge theories containing
Higgs fields \citet{Tong:2008qd}.

With this in mind, it's interesting to reinterpret our results from the point of view of vortices. This is particularly instructive for
SU($2$), where there is no distinction between twisted C-periodic
and twisted periodic boundary conditions for the gauge degrees of
freedom \citet{Kronfeld:1990qu}. The gauge content
of our configurations can therefore be interpreted in terms of twisted
periodic boundary conditions, where the vortex structure is well understood.
Twist in a plane corresponds to an odd number of center vortices
piercing that plane \citet{'tHooft:1979uj}. 

In this case, charge conjugation is carried entirely by the Higgs
field. We will see how a C-periodic/antiperiodic Higgs field modifies the vortex
structure of pure SU($2$) gauge theory and leads to Abelian magnetic
charge. We will then generalise to SU($N$).

First we need a way of locating center vortices, which are generally
thick objects. This proceeds via gauge fixing and center projection. A
common choice in the pure gauge theory is Maximal Center Gauge followed
by a projection of the link variables onto the 'nearest' 
center element \citet{Greensite:2003bk}.
The resultant excitations are thin $\mathbb Z_{N}$ vortices known as P-vortices. These are expected to signal the location of center vortices in the
unprojected configurations. However, since we have a Higgs field at
our disposal it makes more sense to use a modified version of Laplacian
Center Gauge \citet{Vink:1992ys,vanderSijs:1996gn,vanderSijs:1997hi,deForcrand:2000pg}.
After diagonalising $\Phi$ we're left with a residual U(1)$^{N-1}$
gauge symmetry. The idea of Laplacian Center Gauge is to use the lowest-lying
eigenvector of the adjoint Laplacian operator as a faux Higgs field.
We can reduce the gauge symmetry to $\mathbb Z_{N}$ by fixing $N-1$ phases
of this auxialary field. Thin vortices then arise a la Nielsen-Olesen.

We'll follow the construction of de Forcrand and Pepe
\citet{deForcrand:2000pg} which starts from the
adjoint lattice Laplacian, 
\begin{equation}\begin{split}
&-\Delta_{xy}^{AB}({U})\\
&=\sum_{\mu}(2\delta_{x,y}\delta^{AB}-{U}_{\mu}^{AB}(x)\delta_{y,x+\hat{\mu}}-{U}_{\mu}^{BA}(x-\hat{\mu})\delta_{y,x-\hat{\mu}}),\end{split}\label{eq:adjLap}\end{equation}
where $A,B$ are the colour indices, $x,y$ the lattice coordinates, and
${U}_{\mu}^{AB}$  the link variables in the adjoint representation,
\begin{equation}
{U}_{\mu}^{AB}(x)=2\text{Tr}(T^{A}U_{\mu}(x)T^{B}U_{\mu}^{\dagger}(x)).\label{eq:dotU}\end{equation}
Since $\Delta$ is a real symmetric matrix, its eigenvalues are real.
If we take $\lambda_{1}$ to be the smallest eigenvalue, the
corresponding eigenvector 
allows us to associate a real 3-dimensional vector $\phi_{(1)}^{A}(x)$
with each lattice site.
The eigenvalues of $\Delta$ are invariant under gauge transformations
$R(x)$, and $\phi_{(1)}^{A}$ transforms like
an adjoint scalar field \citet{deForcrand:2000pg}, with
$\Phi_{(1)}(x)=\phi_{(1)}^{A}(x)T^{A}$,
\begin{equation}
\Phi_{(1)}(x)\rightarrow  \tilde\Phi_{(1)}(x) =
R^{\dagger}(x)\Phi_{(1)}(x) 
R(x) .
\label{eq:Phitfm}\end{equation}
After diagonalising the physical Higgs field, the transformed field
$\tilde{\Phi}_{(1)}(x)$ will not in general be invariant under
remnant U(1)$^{N-1}$ transformations. The gauge freedom may then
be reduced to $\mathbb Z_{N}$ by eliminating the phases of all
$(N-1)$ sub-diagonal elements. Gauge ambiguities arise when any of the
sub-diagonal elements of $\tilde{\Phi}_{(1)}(x)$ are zero. This
involves two conditions, so gauge ambiguities form lines in three
dimensions. Since they carry quantised center flux, these defects are
identifed as $\mathbb Z_{N}$ vortices \citet{deForcrand:2000pg}.

For SU($2$), note that we have a $\mathbb Z_{2}$ vortex whenever
${\phi}_{(1)}^{A}(x)$  
is parallel or antiparallel to the physical Higgs field in colour space. The relative
sign of ${\phi}_{(1)}^{A}(x)$ and the Higgs gives its
local orientation. In the neighbourhood of a monopole, the Higgs field
has a hedgehog shape in colour space. So there is necessarily some
direction along which ${\phi}_{(1)}^{A}(x)$ and the Higgs
field are collinear. What's more, their relative orientation changes
sign at the location of the Abelian charge. It follows that every monopole
lies on a thin $\mathbb Z_{2}$ vortex, which appears as two oppositely directed strings. Monopoles and anti-monopoles form an alternating bead-like structure on the vortices. See \citet{deForcrand:2000pg} for more details and the generalization to SU($N$).

\begin{figure}
\includegraphics[width=0.95\linewidth]{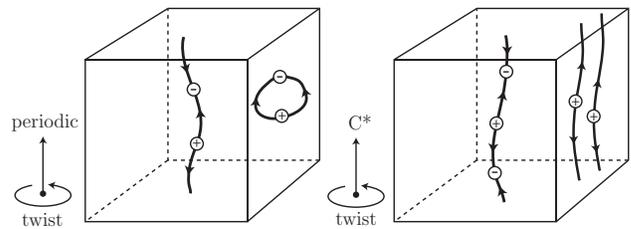} 
\caption{For SU($2$), Abelian monopoles form a bead-like structure on center vortices. To have odd net charge we need an odd number of vortices that contain an odd number of monopoles. i.e. both twist and
charge conjugation.}
\end{figure}

How does this relate to our boundary conditions? Let's start with
SU($2$). Recall that twisted C-periodic and twisted periodic boundary conditions
are equivalent for the gauge links. Twist in a plane forces an odd
number of $\mathbb Z_{2}$ vortices through that plane. For each of
these to contribute an odd number of monopoles/anti-monopoles, the
orientation of $\mathbb Z_{2}$ flux should change an odd
number of times. Therefore ${\phi} ^{A}{\phi}^{A}_{(1)}=2\text{Tr}\Phi\Phi_{(1)}$
should be antiperiodic. The boundary conditions 
\begin{eqnarray}
\Phi(x+L\hat{\jmath}) & = & \pm\Omega_{j}^{\dagger}(x)\Phi(x)\Omega_{j}(x),\nonumber \\
U_{\mu}(x+L\hat{\jmath}) & = & \Omega_{j}^{\dagger}(x)U_{\mu}(x)\Omega_{j}(x+\hat{\mu}),\label{eq:Cbc_2}\end{eqnarray}
 give \begin{equation} \begin{split}
2\text{Tr}&
\Phi(x+N\hat{\jmath})\Phi^{(1)}(x+N\hat{\jmath})  \\ & =\pm2\text{Tr}\Omega_{j}^{\dagger}\Phi(x)\Omega_{j}\Omega_{j}^{\dagger}\Phi^{(1)}(x)\Omega_{j}  \\ & =\pm2\text{Tr}\Phi(x)\Phi^{(1)}(x). \end{split}
\end{equation} 
So if the Higgs field is antiperiodic/C-periodic, $2\text{Tr}\Phi\Phi_{(1)}$ is also antiperiodic, and there will be an odd number of monopoles on every vortex in that direction.

The net magnetic charge is then obtained from simple counting arguments. Closed vortices and vortices through periodic directions do not contribute, since they contain an equal number of monopoles and anti-monopoles. And without twist we can only have an even number of monopoles, since there will be an even number of vortices. For the net charge to be odd, there must be an odd number of directions that are both conjugated and have twist in the orthogonal plane.  We then have an odd number of vortices that contain an odd number of monopoles. This interpretation is in perfect agreement with the results of Sec.~\ref{sub:Allowed-magnetic-charges} and the Appendix.

For the generalisation to SU($N$), it's helpful to start with
a single C-periodic direction.  The main difference now is that
several $\mathbbm Z_N$ vortices are permitted to meet at a point.  We
may have monopole-vortex nets as opposed to the necklaces of
SU($2$). However, as shown in the Appendix and discussed in
Sec.~\ref{subsec:Mixbc}, the center flux through the C-periodic
direction is eliminated for odd $N$ and still restricted to $\mathbbm
Z_2$ for even $N$. This is because the center flux, when viewed as
Abelian flux, must be equal and opposite at the boundary. 

The reversal of flux means that the allowed magnetic charges are governed by
\begin{equation}
\exp{\frac{igQ}{2}}=\exp{i\pi \frac{2m_{\parallel}}{N}},\end{equation} 
where $Q$ is an $N$-vector. The formation of monopole-vortex
nets is reflected in the various solutions for $Q$. The constituent
charges will generally be scattered around the box, connected by vortices
that conserve center flux at each monopole. 

If $2m_{\parallel}/N=1$
then\begin{equation}\begin{split}
Q=\frac{2\pi}{g}(1,1,\dots,1-N)
+\frac{2\pi}{g}2(n_{1},n_{2},\dots,-\sum_{i=1}^{N-1}n_{i}). 
\end{split}\end{equation}
That is, the net always contains an \emph{odd} number of each monopole
species. Of course, this is only possible for even $N$. If $2m_{\parallel}/N=0$
we're left with the second term and hence an \emph{even} number of
each monopole species. Note that this decomposition also applies when all
directions are charge conjugated, since by Eqs.~(\ref{eq:z constraint}) and (\ref{eq:WilsFlux}) we have the same possibilities
for center (and hence Abelian) flux through each half of the box.

\section{Conclusions}

We have shown how twisted C-periodic boundary conditions (\ref{eq:Cbc}) consisting
of complex conjugation and gauge transformations can be used to impose
a non-zero magnetic charge in SU($N$)+adjoint Higgs theory while
preserving translation invariance. This generalises the results obtained 
for SU(2) in Ref.~\citet{Davis:2000kv},
and makes it possible
to study magnetic monopoles in lattice Monte Carlo simulations. In
particular, it will be straightforward to measure the monopole mass
in the same way as in Ref.~\citet{Rajantie:2005hi}.

This method has significant restrictions: It only works for SU($N$)
with even $N$, the charges can only be constrained to be odd or even,
and every residual U(1) group has to have a magnetic charge. Even with 
these restrictions, the method can be used to study quantum monopoles is 
new types of systems, for instance in cases where there are several different types of 
monopoles or an unbroken non-Abelian subgroup. Using methods introduced in Ref.~\cite{Rajantie:2009bk},
one should even be able to find the spectrum of different monopole states, including excited 
states of monopoles.

\begin{acknowledgments}
DM would like to thank the Theoretical Physics Division of the Imperial College London for hospitality and Tanmay Vachaspati for useful correspondence. SE would like to thank Maxim Chernodub for useful correspondence. AR was supported by the STFC and DM was supported by the British Council Researchers' Exchange Programme. LvS gratefully acknowledges support by the Helmholtz International Center for FAIR within the LOEWE program of the State of Hesse.
\end{acknowledgments}
\appendix
%dummy comment inserted by tex2lyx to ensure that this paragraph is not empty

\section{Mixed boundary conditions}

\label{app:mixed}

\subsection{$x$ direction C-periodic, $y,z$ directions periodic}

\label{app:Cpx}
Suppose that we employ boundary conditions with a single C-periodic
direction, chosen to be the $x$ direction. These boundary conditions
may be written as 
\begin{eqnarray}
\Phi(x+L\hat{x})=\Omega_{x}^{\dagger}\Phi^*(x)\Omega_{x},  &&
 U_{\mu}(x+L\hat{x})=\Omega_{x}^{\dagger}U_{\mu}^*(x)\Omega_{x} ,
 \nonumber \\ 
\Phi(x+L\hat{y})=\Omega_{y}^{\dagger}\Phi(x)\Omega_{y}, && 
 U_{\mu}(x+L\hat{y})=\Omega_{y}^{\dagger}U_{\mu}(x)\Omega_{y} ,
 \nonumber \\
\Phi(x+L\hat{z})=\Omega_{z}^{\dagger}\Phi(x)\Omega_{z}, && 
U_{\mu}(x+L\hat{z})=\Omega_{z}^{\dagger}U_{\mu}(x)\Omega_{z} . 
 \nonumber 
\end{eqnarray}
Again assuming constant transition funcitons, 
consistency of the boundary
conditions now requires 
\begin{eqnarray}
\Omega_{x}\Omega_{y} & = & z_{12}\Omega_{y}^*\Omega_{x},\nonumber \\
\Omega_{x}\Omega_{z} & = & z_{13}\Omega_{z}^*\Omega_{x},\nonumber \\
\Omega_{y}\Omega_{z} & = & z_{23}\Omega_{z}\Omega_{y},
\label{eq:C1z}\end{eqnarray}
 where $z_{ij}=e^{i\theta_{ij}} $, $\theta_{ij} = (2\pi/N)\,
 \epsilon_{ijk} m_k $ with $m_k \in \mathbb Z_{N}$, are center elements
as before. Note that charge conjugation only ever happens on one side
of the equation.

\begin{figure}[t]
\includegraphics[width=\linewidth]{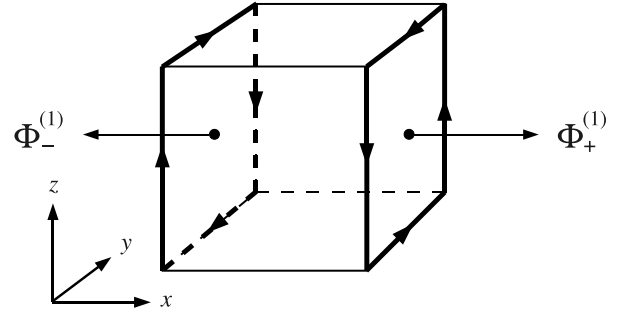}

\caption{Quantised Abelian fluxes of equal strength in opposite
  directions, $\Phi^{(1)}_+ = \Phi^{(1)}_-$, with C-periodic
  $x$ direction. \label{fig:1dFluxes}}

\end{figure}

The gauge transformations to diagonalise the Higgs field have the
following genuine boundary conditions,
\begin{eqnarray}
R( x + L \hat x) &=& \Omega^\dagger_x R^*( x) , \\ 
R( x + L \hat y) &=& \Omega^\dagger_y R( x) , \nonumber\\
R( x + L \hat z) &=& \Omega^\dagger_z R( x) , \nonumber
\end{eqnarray}
and we define the following doubly translated $R$'s at the far edges
and corner by lexicographic order,
\begin{eqnarray}
  R(L,L,r) &\equiv & \Omega_y^\dagger \Omega_x^\dagger R^*(0,0,r) ,\\
  R(L,r,L) &\equiv & \Omega_z^\dagger \Omega_x^\dagger R^*(0,r,0)
  ,\nonumber \\
  R(r,L,L) &\equiv & \Omega_z^\dagger \Omega_y^\dagger R(r,0,0) ,\nonumber
\end{eqnarray}
for $r = 0, \dots L-1$, and
\begin{equation} 
  R(L,L,L) \equiv \Omega^\dagger_z \Omega^\dagger_y \Omega^\dagger_x
  R^*(0,0,0) .
 \end{equation}
It is straightforward to derive the corresponding boundary conditions
for the Abelian projected fields~(\ref{eq:kronalpha}), 
\begin{eqnarray}
\alpha_{i}^{a}(x+L\hat{x}) & = & - \alpha_{i}^{a}(x),\nonumber \\
\alpha_{i}^{a}(x+L\hat{y}) & = & \alpha_{i}^{a}(x),\nonumber \\
\alpha_{i}^{a}(x+L\hat{z}) & = & \alpha_{i}^{a}(x),
\end{eqnarray}
with the following exceptions,
\begin{eqnarray}
\alpha_{y}^{a}(L,L-1,r) & = & - \alpha_{y}^{a}(0,L-1,r)
-\smallfrac{2\pi}{N} \, m_3  , \nonumber \\
\alpha_{z}^{a}(L,r,L-1) & = & - \alpha_{z}^{a}(0,r,L-1 )
+\smallfrac{2\pi}{N} \, m_2  , \nonumber \\
\alpha_{z}^{a}(r,L,L-1) & = &  \alpha_{z}^{a}(r,0,L-1) 
-\smallfrac{2\pi}{N} \, m_1  ,
\end{eqnarray}
$r = 0, \dots L-1$, and
\begin{eqnarray}
\alpha_{y}^{a}(L,L-1,L) & = & - \alpha_{y}^{a}(0,L-1,0)
-\smallfrac{2\pi}{N} \, m_3   \nonumber \\
 & = & - \alpha_{y}^{a}(0,L-1,L)
-\smallfrac{2\pi}{N} \, m_3 \nonumber \\
 & = & \alpha_{y}^{a}(L,L-1,0) ,
\end{eqnarray}
as well as
\begin{eqnarray}
\alpha_{z}^{a}(L,L,L-1) & = & - \alpha_{z}^{a}(0,0,L-1) 
-\smallfrac{2\pi}{N} \, (m_1 - m_2)  \nonumber\\
 & = & - \alpha_{z}^{a}(0,L,L-1) 
-\smallfrac{2\pi}{N} \, (2m_1 - m_2)  \nonumber\\
 & = & \alpha_{z}^{a}(L,0,L-1) 
-\smallfrac{2\pi}{N} \, m_1  .
\end{eqnarray}
As expected, it follows that the Abelian field strengths
$\alpha_{ij}^{a}(x)$ (\ref{eq:plaqangles}) are periodic in
the $y$ and $z$ directions, but anti-periodic in the $x$ direction, again
with one exception. And that exception is
\begin{equation}
 \alpha_{23}(L,L-1,L-1) =  - \alpha_{23}(0,L-1,L-1) -
 \smallfrac{2\pi}{N} \, 2 m_1 .  
\end{equation} 
It is this single plaquette where our single-valued gauge
transformation $R( x)$ has moved the net magnetic flux to.
It leads to opposite fluxes each of strength $(2\pi/gN)  m_1$ through the
faces at $x=0$ and $L$ as illustrated in Figure~\ref{fig:1dFluxes}. For
the total flux along the positive $x$ 
direction, for example, we obtain
\begin{eqnarray} 
  \Phi_+^{(1)} &=& - \smallfrac{1}{g} \sum_{r=0}^{L-1} \big( 
             \alpha_y(L,r,0) + \alpha_z(L,L,r) \nonumber\\
&& \hskip 2cm  - \alpha_y(L,r,L) - \alpha_z (L,0,r) \big) \nonumber\\
&=& \smallfrac{1}{g} \smallfrac{2\pi}{N} \, m_1 .  \label{eq:Phi1p}
\end{eqnarray}
Analogously, we obtain for the total flux in the negative $x$ direction at
$x=0$,
\begin{eqnarray} 
  \Phi_-^{(1)} &=& - \smallfrac{1}{g} \big( \alpha_z (0,0,L-1) -
  \alpha_z(0,L,L-1) \big)  \nonumber\\
 &=&   \smallfrac{1}{g} \smallfrac{2\pi}{N} \, m_1 \, = \Phi^{(1)}_+ .
\label{eq:Phi1m}
\end{eqnarray}

\noindent
The analogous Abelian projected fluxes in the $y$ and $z$ directions are not
quantised, because they each involve anti-periodic line segments, but
they are both conserved. We have
\begin{eqnarray} 
 \Phi_+^{(2)} &=& -\smallfrac{1}{g} \left(2\sum_{r=0}^{L-1} \label{eq:Phi2}
   \alpha_z(0,0,r)  - \smallfrac{2\pi}{N} \, m_2 \right) , \nonumber\\
   &= &   - \Phi_-^{(2)}  
 \end{eqnarray}
 and
\begin{eqnarray} 
 \Phi_+^{(3)} &=& \smallfrac{1}{g} \left(2\sum_{r=0}^{L-1} \label{eq:Phi3}
   \alpha_y(0,r,0)  + \smallfrac{2\pi}{N} \, m_3 \right) , \nonumber\\
   &= &   - \Phi_-^{(3)} 
 \end{eqnarray}
Therefore, the Abelian projected fluxes in the $y$ and $z$ directions
are not quantised but they are conserved, i.e.
\begin{equation} 
         \Phi_+^{(2)} + \Phi_-^{(2)} =
         \Phi_+^{(3)} + \Phi_-^{(3)} = 0 .
\end{equation}
There is one extra source of strength $2 m_1/N $ in units of magnetic charge
$(2\pi/g)$ whose entire flux goes along the $x$ direction through
the $\alpha_{23} $ plaquettes in the opposite $y=z=L-1$ corners at
$x=0$ and $L$, 
\begin{equation}
  Q  =  \Phi_+^{(1)} + \Phi_-^{(1)} = \smallfrac{2\pi}{g} \,
  \smallfrac{2m_1}{N} \; , \label{eq:charge1Cp}
\end{equation}
again modulo $(4\pi/g) $ and the same for all $a = 1, \dots N-1$.

It would be interesting if the twist angle in the plane perpendicular
to the C-periodic direction were permitted to be a phase other than 0
or $\pi$. Unfortunately this is not the case. The proof involves
permutations of the twist matrices as before. With C-periodic b.c.'s
in the $x$ direction, comparison of
\begin{equation}
\begin{split}
\Omega_{x}\Omega_{y}\Omega_{z} &= z_{23}
                       \Omega_{x}\Omega_{z}\Omega_{y} \\
 &= z_{23}z_{13}\Omega_{z}^*\Omega_{x}\Omega_{y}\\
 &= z_{23}z_{13}z_{12}\Omega_{z}^*\Omega_{y}^*\Omega_{x} 
\end{split}
\label{eq:permute1c}
\end{equation}
 with
\begin{equation}
\begin{split}
\Omega_{x}\Omega_{y}\Omega_{z} &= z_{12}
                       \Omega_{y}^*\Omega_{x}\Omega_{z} \\
 &= z_{12}z_{13}\Omega_{y}^*\Omega_{z}^*\Omega_{x}\\
 &= z_{12}z_{13}z_{32}\Omega_{z}^*\Omega_{y}^*\Omega_{x} 
\end{split}
\label{eq:permute2c}
\end{equation}
 yields
\begin{equation}
  z_{23} = z_{32}= z_{23}^* .
\end{equation}
 Therefore\begin{equation}
m_1 =\begin{cases}
0 & \text{for odd } N\\
0\;\text{or}\; N/2 & \text{for even } N .
\end{cases}\end{equation}
It follows that the allowed charges are exactly those found for
fully C-periodic boundary conditions.

\begin{figure}[t]
\includegraphics[width=0.9\linewidth]{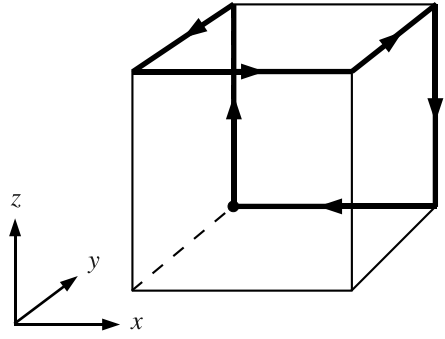}

\caption{Integration curve for C-periodic  $y$ and $z$ directions. 
\label{fig:mixed curve}}

\end{figure}

\subsection{$y,z$ directions C-periodic, $x$ direction periodic}

We can also consider boundary conditions with two C-periodic directions,
chosen to be the y and z directions. Then the consistency conditions
are modified to \begin{eqnarray}
\Omega_{x}^{*}\Omega_{y} & = & z_{12}\Omega_{y}\Omega_{x},\nonumber \\
\Omega_{x}^{*}\Omega_{z} & = & z_{13}\Omega_{z}\Omega_{x},\nonumber \\
\Omega_{y}^{*}\Omega_{z} & = & z_{23}\Omega_{z}^{*}\Omega_{y}\end{eqnarray}
 with $z_{ij}=e^{i\theta_{ij}}\in \mathbb Z_{N}$,  $\theta_{ij} = (2\pi/N)\,
 \epsilon_{ijk} m_k $. The Abelian projected fields
inherit boundary conditions with anti-periodicity in both C-periodic
directions,\begin{eqnarray}
\alpha_{i}(x+L\hat{x}) & = & \alpha_{i}(x),\nonumber \\
\alpha_{i}(x+L\hat{y}) & = & -\alpha_{i}(x),\nonumber \\
\alpha_{i}(x+L\hat{z}) & = & -\alpha_{i}(x),\label{eq:simple b.c.'s 2 c*}\end{eqnarray}
 except for the special cases, which in this case are,
\begin{eqnarray}
\alpha_{y}^{a}(L,L-1,r) & = &  \alpha_{y}^{a}(0,L-1,r)
-\smallfrac{2\pi}{N} \, m_3  ,  \label{eq:tagged b.c.'s 2 c*}\\
\alpha_{z}^{a}(L,r,L-1) & = &  \alpha_{z}^{a}(0,r,L-1 )
+\smallfrac{2\pi}{N} \, m_2  , \nonumber \\
\alpha_{z}^{a}(r,L,L-1) & = &  - \alpha_{z}^{a}(r,0,L-1) 
-\smallfrac{2\pi}{N} \, m_1  , \nonumber
\end{eqnarray}
for $r = 0, \dots L-1$, and
\begin{eqnarray}
\alpha_{y}^{a}(L,L-1,L) & = & - \alpha_{y}^{a}(0,L-1,0)
+\smallfrac{2\pi}{N} \, m_3   \nonumber \\
 & = &  \alpha_{y}^{a}(0,L-1,L)
+\smallfrac{2\pi}{N} \, m_3 \nonumber \\
 & = & - \alpha_{y}^{a}(L,L-1,0) ,  
\end{eqnarray}
as well as
\begin{eqnarray}
\alpha_{z}^{a}(L,L,L-1) & = & - \alpha_{z}^{a}(0,0,L-1) 
-\smallfrac{2\pi}{N} \, (m_1 + m_2)  \nonumber\\
 & = & \alpha_{z}^{a}(0,L,L-1) 
-\smallfrac{2\pi}{N} \, m_2  \nonumber\\
 & = & - \alpha_{z}^{a}(L,0,L-1) 
-\smallfrac{2\pi}{N} \, m_1  .
\end{eqnarray}
To find the total flux we integrate $\alpha_{i}^{a}(x)$
around the curve shown on the right of Figure \ref{fig:mixed curve},
and double the result,
\begin{eqnarray}
Q &=& -\frac{2}{g} \sum_{r=0}^{L-1} \Big( \alpha_z(0,L,r) -
\alpha_y(0,r,L) + \alpha_x(r,0,L) \nonumber \\
&& \hskip .5cm 
 - \alpha_z(L,L,r) +\alpha_y(L,r,L) - \alpha_x(r,L,0) \Big) .  \nonumber
\end{eqnarray} 
The $x$ links here are translated in two anti-periodic directions
relative to one another and hence cancel. The $y$ and $z$ links are
related to one another by a single periodic translation along the $x$
direction and therefore also cancel except for contributions from
the special links above, which yield
\begin{eqnarray}
Q & = & - \frac{2}{g}\Big( \alpha_y(L,L-1,L) - \alpha_y(0,L-1,L) \\
  && \hskip .5 cm - \alpha_z(L,L,L-1) + \alpha_z(0,L,L-1) 
      \Big) \nonumber \\
 & = & \smallfrac{2\pi}{g} \smallfrac{2}{N} \big(m_{2}+m_{3}\big) ,
\end{eqnarray}
modulo $(4\pi/g) $ and the same for all $a = 1, \dots N-1$, as before.
And as before, we find that the center fluxes in the C-periodic
directions are restricted. Comparison of
\begin{equation}
\begin{split}\Omega_{x}^{*}\Omega_{y}\Omega_{z}^{*} &=
     z_{32}\Omega_{x}^{*}\Omega_{z}\Omega_{y}^{*} \\
  & =z_{32}z_{13}\Omega_{z}\Omega_{x}\Omega_{y}^{*} \\
 & =z_{32}z_{13}z_{21}\Omega_{z}\Omega_{y}^{*}\Omega_{x}^{*}
\end{split}
\label{eq:permute1cc}
\end{equation}
with
\begin{equation}
\begin{split}\Omega_{x}^{*}\Omega_{y}\Omega_{z}^{*}
 & =z_{12}\Omega_{y}\Omega_{x}\Omega_{z}^{*} \\
 & =z_{12}z_{31}\Omega_{y}\Omega_{z}^{*}\Omega_{x}^{*} \\
 & =z_{12}z_{31}z_{32}\Omega_{z}\Omega_{y}^{*}\Omega_{x}^{*}
\end{split}
\label{eq:permute2cc}\end{equation}
now yields
\begin{equation}
z_{21}^{2}z_{13}^{2}=1.
\end{equation}
So, 
\begin{equation}
m_2 + m_3  =
\begin{cases}
0 & \text{for odd } N\\
0\;\text{or}\; N/2 & \text{for even } N .
\end{cases}
\end{equation}
Once more, we are left with the same possibilities for the Abelian
magnetic charges. We conclude that the allowed charges are identical
whether we have one, two, or all three directions charge conjugated.

\vfill

\bibliographystyle{apsrev-Lo}
\bibliography{monopolebc}

\end{document}